\documentclass[conference]{IEEEtran}
\IEEEoverridecommandlockouts
\usepackage{cite}
\usepackage{amsmath,amssymb,amsfonts}
\usepackage{algorithmic}
\usepackage{graphicx}
\usepackage{xcolor}
\def\BibTeX{{\rm B\kern-.05em{\sc i\kern-.025em b}\kern-.08em
    T\kern-.1667em\lower.7ex\hbox{E}\kern-.125emX}}

\usepackage[utf8]{inputenc}
\usepackage{algorithm}
\usepackage[subrefformat=parens]{subcaption}
\usepackage{comment}
\usepackage{siunitx}
\usepackage{url}
\usepackage[bookmarks=false]{hyperref}
\usepackage{textcomp}

\makeatletter
\newcommand{\algorithmicfunc}{\textbf{function}}
\newcommand{\algorithmicendfunc}{\algorithmicend\ \algorithmicfunc}

\newenvironment{ALC@func}{\begin{ALC@g}}{\end{ALC@g}}
\newcommand{\FUNC}[2][default]{\ALC@it\textbf{function}\ #2%
\ALC@com{#1}\begin{ALC@func}}
\ifthenelse{\boolean{ALC@noend}}{
    \newcommand{\ENDFUNC}{\end{ALC@func}}
  }{
    \newcommand{\ENDFUNC}{\end{ALC@func}\ALC@it\algorithmicendfunc}
  } 
\makeatother

\IEEEoverridecommandlockouts
\usepackage{tikz}
\usepackage{lipsum}

\newcommand\copyrighttext{%
  \footnotesize \textcopyright 2020 IEEE. Personal use of this material is permitted.
  Permission from IEEE must be obtained for all other uses, in any current or future
  media, including reprinting/republishing this material for advertising or promotional
  purposes, creating new collective works, for resale or redistribution to servers or
  lists, or reuse of any copyrighted component of this work in other works.
  DOI: \href{https://ieeexplore.ieee.org/document/9202766}{10.1109/COMPSAC48688.2020.00-44}}
\newcommand\copyrightnotice{%
\begin{tikzpicture}[remember picture,overlay]
\node[anchor=south,yshift=10pt] at (current page.south) {\fbox{\parbox{\dimexpr\textwidth-\fboxsep-\fboxrule\relax}{\copyrighttext}}};
\end{tikzpicture}%
}

\begin{document}

\title{Divider: Delay-Time Based Sender Identification in Automotive Networks}

\author{\IEEEauthorblockN{Shuji Ohira, Araya Kibrom Desta}
\IEEEauthorblockA{\textit{Graduate School of Science and Technology} \\
\textit{Nara Institute of Science and Technology}\\
Nara, Japan \\
\{ohira.shuji.ok2, kibrom\_desta.araya.js3\}@is.naist.jp}
\and
\IEEEauthorblockN{Tomoya Kitagawa}
\IEEEauthorblockA{
jptomoya@gmail.com}
\and
\IEEEauthorblockN{Ismail Arai, Kazutoshi Fujikawa}
\IEEEauthorblockA{\textit{Information Initiative Center} \\
\textit{Nara Institute of Science and Technology}\\
Nara, Japan \\
\{ismail, fujikawa\}@itc.naist.jp
}
}
\maketitle
\copyrightnotice

\begin{abstract}
Controller Area Network (CAN) is one of the in-vehicle network protocols that is used to communicate among Electronic Control Units (ECUs) and has been de-facto standard.
CAN is simple and has several vulnerabilities such as unable to distinguish spoofing messages because it doesn't support any authentication or sender identification properties. 
In previous work, some voltage-based methods to identify the sender node have been proposed.
The methods can identify ECUs with high accuracy. 
However, the accuracy of source identification depends on a feature that is extracted from a continuous function of voltage use sampling.
In general, as the sampling rate increases, the accuracy of identification is improved.
Though the amount of data used for the identification increases too.
Hence, it is desired to create an Intrusion Detection System (IDS) that identifies ECUs using few sampling features as there is a limited computing resource in vehicles. 
In this paper, we propose a delay-time based sender identification method of ECUs. 
We confirm that the proposed method achieved a true positive rate of 96.7\% in CAN bus prototype against spoofing attack from a compromised ECU, detecting spoofing attack from an unmonitored ECU with a true positive rate of 98.0\% in real-vehicle.
\end{abstract}


\section{Introduction}
Due to the increase in the number of automobiles that connect to the internet, cyberattacks on automobiles are becoming a severe problem \cite{nie2017free,miller2015remote}. 
These attacks abuse vulnerable Controller Area Network (CAN) \cite{can20pdf98:online} which is one of the in-vehicle network protocols that is used to communicate among Electronic Control Units (ECUs) and has been de-facto standard. 
Nie et al. successfully controlled some automotive functions, exploiting the vulnerabilities in a CAN and a browser in the in-vehicle system implemented by WebKit of the old version \cite{nie2017free}.
Therefore, cybersecurity countermeasures for automobiles are urgently required. 

Countermeasures such as encryption and authentication have been proposed to prevent spoofing, sniffing and replay attacks. 
Since CAN has only a short data field of 8 bytes and limited bandwidth, adding a Message Authentication Code (MAC) is not practical. Moreover, since some authentication methods \cite{van2011canauth,groza2012libra} require pre-shared keys and does not concern itself with key exchanges. Therefore, these methods are impractical in automobiles that are already widespread.

While the Intrusion Detection System (IDS) has a good advantage in terms of effectiveness and high compatibility in automotive security different from encryption and authentication. One such case is, IDSs based on characteristics of digital-level (e.g. frequency, entropy, ID sequence). The approach can be adapted easily to the CAN bus of modern automobiles. However, these approaches typically have higher false positives for some attack types. For instance, ID sequence-based IDS \cite{marchetti2017id} cannot detect replay attacks, in which an adversary sends messages of the same ID sequence. Hence, we should consider an IDS that can detect various attack types.

An IDS based on physical-level features such as the voltage has been proposed. 
The methods based on voltage use result of sampling continuous function as features. Thus, the accuracy of identification depends on the sampling rate. In general, as the sampling rate increases, the accuracy of identification is improved \cite{kneib2018scission}. However, the amount of data used for the identification increases too.
Hence, it is desired to create an IDS that identifies ECUs using few sampling features in vehicles limited computing resources. Therefore, we focus on identifiable characteristics with few sampling.
In this research, we propose a delay-time based sender identification method. 
The proposed method can identify the ECUs with a sampling count less than the voltage based because the delay-time is observed only from each rising edge of the CAN message. 

The main contributions of this study can be summarized as follows:
\begin{enumerate}
  \item We propose delay-time based sender identification method called Divider. Our method uses new characteristics in the identification of ECUs. Divider does not use continuous characteristics such as voltage, but the delay-time to be observed in each rising edge of the CAN message. 
  Hence, Divider can identify the ECUs with a sampling count less than the voltage based method. Besides, the delay-time can be observed at only one probe point. 

  \item Divider achieved a true positive rate of \SI{96.7}{\%} in CAN bus prototype against spoofing attack from a compromised ECU and a true positive rate of \SI{98.0}{\%} in real-vehicle against spoofing attack from an unmonitored ECU.
\end{enumerate}

\section{Controller Area Network}
\label{section:background}

\begin{figure}[tbp]
  \centering
  \begin{minipage}{0.7\hsize}
   \centering
   \includegraphics[width=1.00\textwidth, height=0.3\hsize]{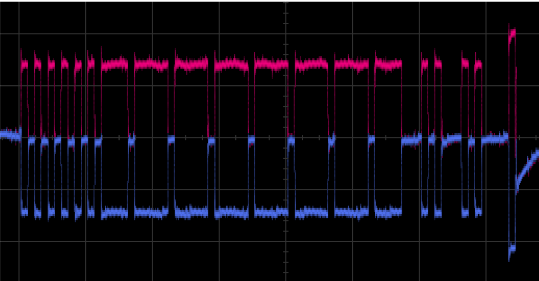}
   \subcaption{An example of CAN signal.}\label{1_a}
  \end{minipage}
  \begin{minipage}{1.00\hsize}
   \centering
   \includegraphics[width=1.00\textwidth]{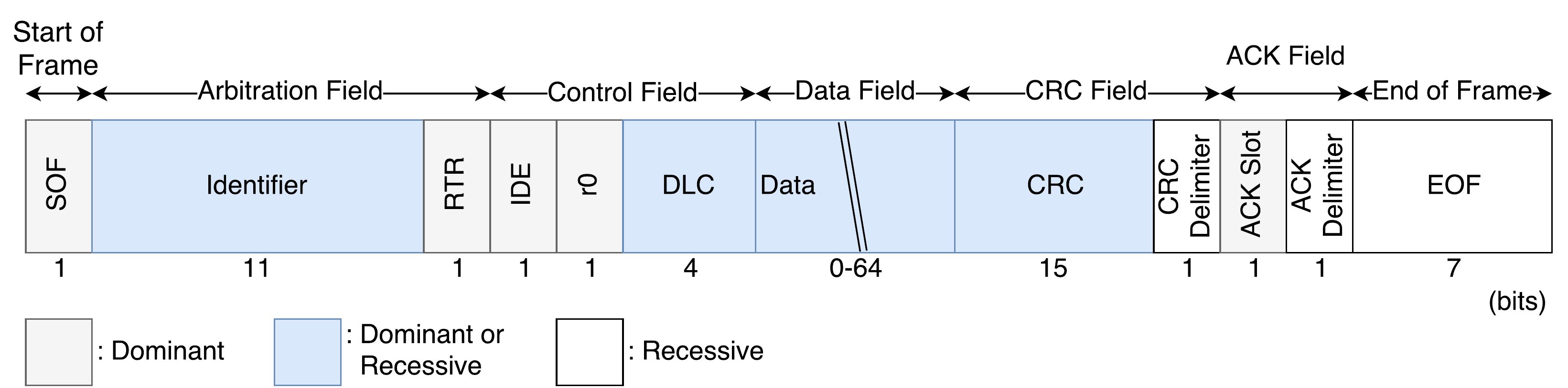}
   \subcaption{CAN data frame format.}\label{1_b}
  \end{minipage}
   \caption{CAN data frame.}
   \label{fig:CAN_frame}
\end{figure}

CAN is one of the in-vehicle network protocols that is widely used to communicate among ECUs and has been a de-facto standard.
Typical CAN node consists of
Micro Controller Unit (MCU), CAN controller and CAN transceiver.
The CAN controller processes various frames according to the CAN protocol.
The CAN transceiver converts the logical level (low and high) and the CAN bus level (dominant and recessive) between the CAN bus and the CAN controller.
ISO 11898 gives the High-Speed CAN bus specification.
The specifications are given for the maximum baud rate of \SI{1}{\mega bps} and a maximum bus length of \SI{40}{\meter} with up to 30 nodes can be connected.
A twisted-pair cable is used to ensure robust noise immunity on the CAN bus.
The two wires are called CAN-L and CAN-H respectively.
If dominant (logical 0) is transmitted, CAN-H is driven towards higher voltage (typically \SI{3.5}{V}) and CAN-L is driven towards lower voltage \SI{1.5}{V}, but when the recessive (logical 1) is transmitted, both CAN-H and CAN-L become \SI{2.5}{V}, as Fig. \ref{fig:CAN_frame} \subref{1_a}. In Fig. \ref{fig:CAN_frame} \subref{1_a}, recessive is inserted in a fixed period due to a bit called stuff-bit which is inserted in after succeeding 5 bit of same logic for synchronization.
Also, the CAN bus is terminated at both ends with \SI{120}{\Omega} resistors to prevent signal reflections.

As Fig. \ref{fig:CAN_frame} \subref{1_b} shows, a CAN data frame does not contain a field that indicate its sender.
Hence, a receiver cannot distinguish which ECU transmitted a CAN message.
The CAN is simple and has several vulnerabilities such as unable to distinguish spoofing messages due to no authentication.
An adversary can change the speedometer reading, unlock the door, turn on the light and so on by sending a malicious message on the CAN bus.

\section{Related works}
\label{section:related}
\subsection{IDS based on characteristics of digital-level}

In \cite{song2016intrusion} frequency-based intrusion detection method has been proposed. However, there are limitations to the kinds of attacks this method can detect. For example, the frequency-based IDS is not able to detect a message mimicking the original message's frequency. In the same catagory, an IDS based the message ID sequence has also been proposed \cite{marchetti2017id}. However, an adversary may inject malicious messages under legitimate message ID sequences such as replay attacks.

The entropy of arbitration ID in fixed interval-based IDSs have been proposed \cite{muter2011entropy,marchetti2016evaluation,wu2018sliding}. In these methods, if an adversary injects one fake message per fixed interval, these IDSs cannot detect the attack because the entropy of the interval is almost the same value as usual. To detect the attack, these IDSs must make the fixed interval a small value. Then, the false positive rate of these IDSs will increase due to enormous influence per one message against the entropy.

The IDS on a CAN using deep learning has been proposed \cite{taylor2016lstm,kang2016intrusion}. These methods cannot be realized on in-vehicle computers in restricted resources.

Although digital level attack detection has been extensively studied, IDS based on characteristics of the digital-level has some limitations. It is necessary to consider more advanced attack detection methods.

\subsection{IDS based on characteristics of physical-level}
Murvay et al. firstly proposed a method for sender identification using physical characteristics in CAN \cite{murvay2014source}.  
Choi et al. proposed an improved version of Murvay's method \cite{choi2018identifying}. 
They embed a fixed bit string into the extended identifier field of the CAN frame and sample the signal and identify ECUs by using 17 different features. Hence, these methods cannot be implemented on the normal CAN because they require the extended frame format in CAN. 

Cho et al. proposed a system for identifying an attacker by using voltage difference among ECUs called Viden \cite{cho2017viden}.
They implemented the system on MCU of a lower sampling rate (\SI{50}{kS/s}) than CAN bus bit rate. Therefore, Viden requires 2-3 messages to output a voltage instance and updates the profiles.
Thus the first forged message will be accepted.

Another approach called Clock-based IDS (CIDS) to identify the sender node has been proposed by Cho et al \cite{cho2016fingerprinting}.
Although CIDS does not need special hardware, it does not apply to non-periodic messages and results in lowering identification accuracy on such messages.

Besides, since Viden and CIDS rely on multiple messages to make detection and identification, these methods have vulnerability against the Hill-climbing-style attack \cite{foruhandeh2019simple}, in which an attacker sends gradually malicious messages without being either detected or identified. To be robust against the Hill-climbing-style attack, IDS has to detect the attacks using features acquired in one message \cite{choi2018identifying,kneib2018scission,foruhandeh2019simple}.

Scission \cite{kneib2018scission} improved a problem in the sender identification method proposed by Choi et al \cite{choi2018identifying}, which the method could not get significant characteristics such as the overshoot. As a result, Scission achieves higher accuracy \SI{99.85}{\%} of identification than Viden and the method of Choi et al. 
However, since Scission uses Fourier Transform to calculate the features of the frequency domain, the time complexity of Scission is $\mathrm{\Omega}(n\log n)$ which is higher than the time complexity of SIMPLE \cite{foruhandeh2019simple}. 
Because SIMPLE only use mean of voltage as feature of ECUs, time complexity is $\mathrm{\Theta}(n)$. In addition, since these sender identification methods use result of sampling continuous function, the accuracy of identification depends on the sampling rate. In general, as the sampling rate increases, the accuracy of identification is improved. But the amount of data used for the identification increase too.
Hence, IDS which is limited in computing resources on the in-vehicle system needs to be able to identify ECUs with few sampling. Therefore, we focus on the identification method using another characteristics with few sampling.

\section{Delay-time based sender identification: Divider}
\label{section:proposed}
In the following section, we introduce the Divider, the proposed method, to identify sender ECUs in CAN bus. First, we describe the adversarial models that Divider assumes. Next, we introduce the proposed method. Finally, we describe the implementation method of Divider. 
\subsection{Adversarial models}
In our research, we have considered two types of adversarial models. These models are based on an example considered by some researchers \cite{checkoway2011comprehensive} and the actual attack against Jeep Cherokee \cite{miller2015remote}. 
\subsubsection{Compromised ECU}

The first model is an ECU exploited by adversary through attack surfaces such as Wi-Fi or Bluetooth. Since the ECU has some connectivity interfaces such as Wi-Fi or Bluetooth, the adversary may exploit the attack surfaces \cite{checkoway2011comprehensive}. However, because the ECU communicates the other ECUs using CAN, Divider can monitor its messages. Hence, Divider can detect adversary's illegal ID attack from a compromised ECU. 
\subsubsection{Unmonitored ECU}

The second model is based on the hacking of Jeep Cherokee \cite{miller2015remote}, and the adversary from OBD-II port. In actual hacking of Jeep Cherokee, Miller and Valasek exploited a passive or unmonitored ECU's update mechanism to inject their code. As a result, the ECU unmonitored by IDS can attack CAN. Also, some researchers attach the OBD-II port to analyze and log the messages on the CAN bus. Thus, we have to detect this adversarial model too. In other words, we must suppose the attack from an ECU of which Divider does not learn features.

\subsection{Framework of Divider}
In this section, we introduce a framework of Divider. Divider is mainly organized by three phases (data acquisition, feature extraction, classification). In the first phase of data acquisition, we collect the delay-time from each ECU. In other words, the data acquisition phase converts analog information to a digital value. Next, in the feature extraction phase, Divider extracts some statistical characteristics from delay-time which are collected in the previous phase. Finally, Divider classifies the statistical characteristics of CAN message using a classification algorithm. In the following sections, we explain each phase in order. 

\subsection{Data acquisition}
\subsubsection{Definition of delay-time in CAN}
\label{section:delay}
\begin{figure}[tbp]
  \centering
   \includegraphics[width=0.25\textwidth]{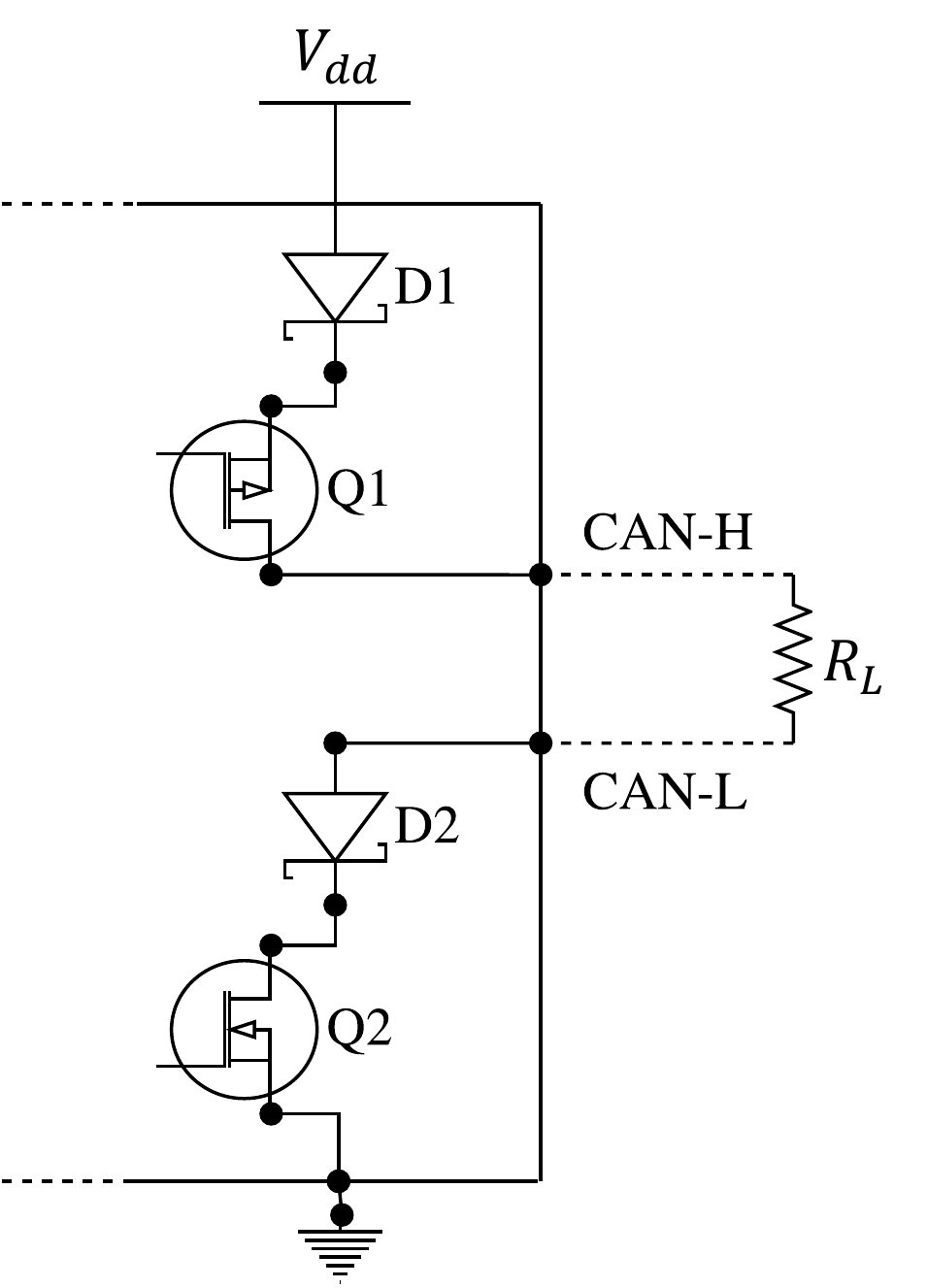}
   \caption{CAN transceiver equivalent.}
   \label{fig:trans_equivalent}
\end{figure}

In this section, we introduce the definition of delay-time in the proposed method.
Fig. \ref{fig:trans_equivalent} shows output schematic of typical CAN transceiver \cite{Ti2016}.
The upper side output structure consists of a series diode (D1) and an N-channel FET (Q1).
The lower side output structure consists of a series diode (D2) and a P-channel FET (Q2).
The upper side diode (D1) prevents reverse current flow to $V_{dd}$ while the voltage on the CAN-H pin rises above $V_{dd}$.
$R_L$ indicates the load resistor.
$R_L$ equals $\SI{60}{\Omega}$ as parallel combined resistance of two terminated $\SI{120}{\Omega}$ resistors on High-Speed CAN bus.
CAN-L and CAN-H are weakly biased to \SI{2.5}{V} during the recessive state.
The delay-time of the signal level transitions is generally determined by the switching time of the transistor and the time during which the load capacitance of the output is charged and discharged.
Factors of the load capacitance include three types of output capacitance at the gate of the transistor, input capacitance of the gate and wiring capacitance.
The factors of these delay-times are different for each CAN node.
The main idea is to use this fact to identify the sender ECU of CAN message.

\begin{figure}[tbp]
  \centering
  \begin{minipage}{0.7\hsize}
   \centering
   \includegraphics[width=1.00\textwidth]{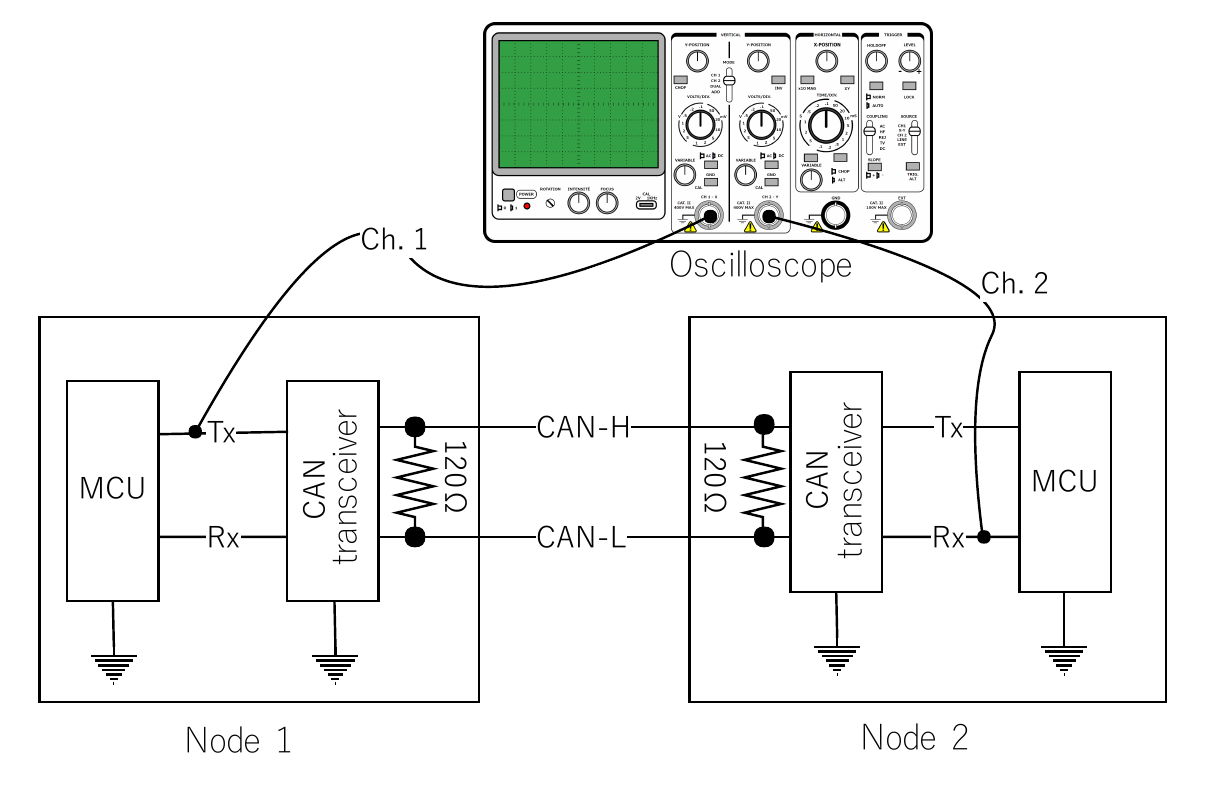}
   \subcaption{Environment to measure delay.}\label{2_a}
  \end{minipage}
  \begin{minipage}{0.7\hsize}
   \centering
   \includegraphics[width=1.00\textwidth]{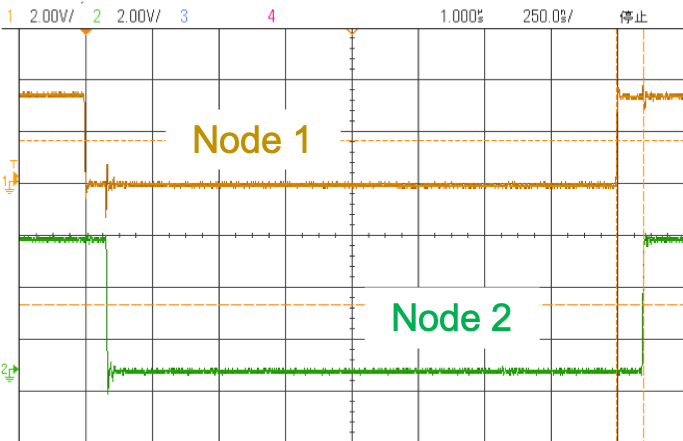}
   \subcaption{Delay between Node 1 (sender) and Node 2 (receiver).}\label{2_b}
  \end{minipage}
   \caption{Measurement of delay-time.}
   \label{fig:trans_delay}
\end{figure}

We experimented to observe these delays in the actual environment.
The experimental environment and the delay-time in the environment are shown in Fig. \ref{fig:trans_delay} \subref{2_a}. 
As shown in Fig. \ref{fig:trans_delay} \subref{2_a}, we constructed the environment from two ECUs and we observed Tx (Node 1) and Rx (Node 2) with a oscilloscope.
Fig. \ref{fig:trans_delay} \subref{2_b} shows a delay between Node 1 (upper in Fig. \ref{fig:trans_delay} \subref{2_b}) and Node 2 (lower in Fig. \ref{fig:trans_delay} \subref{2_b}).
Also, the maximum sampling rate of the oscilloscope is \SI{2.0}{GS/s}. Therefore, the time resolution is $10^{12}/(2.0 \times 10^{9}) = \SI{500}{ps}.$
The fall delay-time of Tx to Rx is \SI{82.0}{ns}, and the reverse is \SI{99.0}{ns} in the example given in Fig. \ref{fig:trans_delay} \subref{2_b}. The difference between rise and fall is \SI{17}{ns}. Divider uses this difference between ECUs to distinguish each node.

\begin{figure}[tbp]
  \centering
   \includegraphics[width=0.47\textwidth]{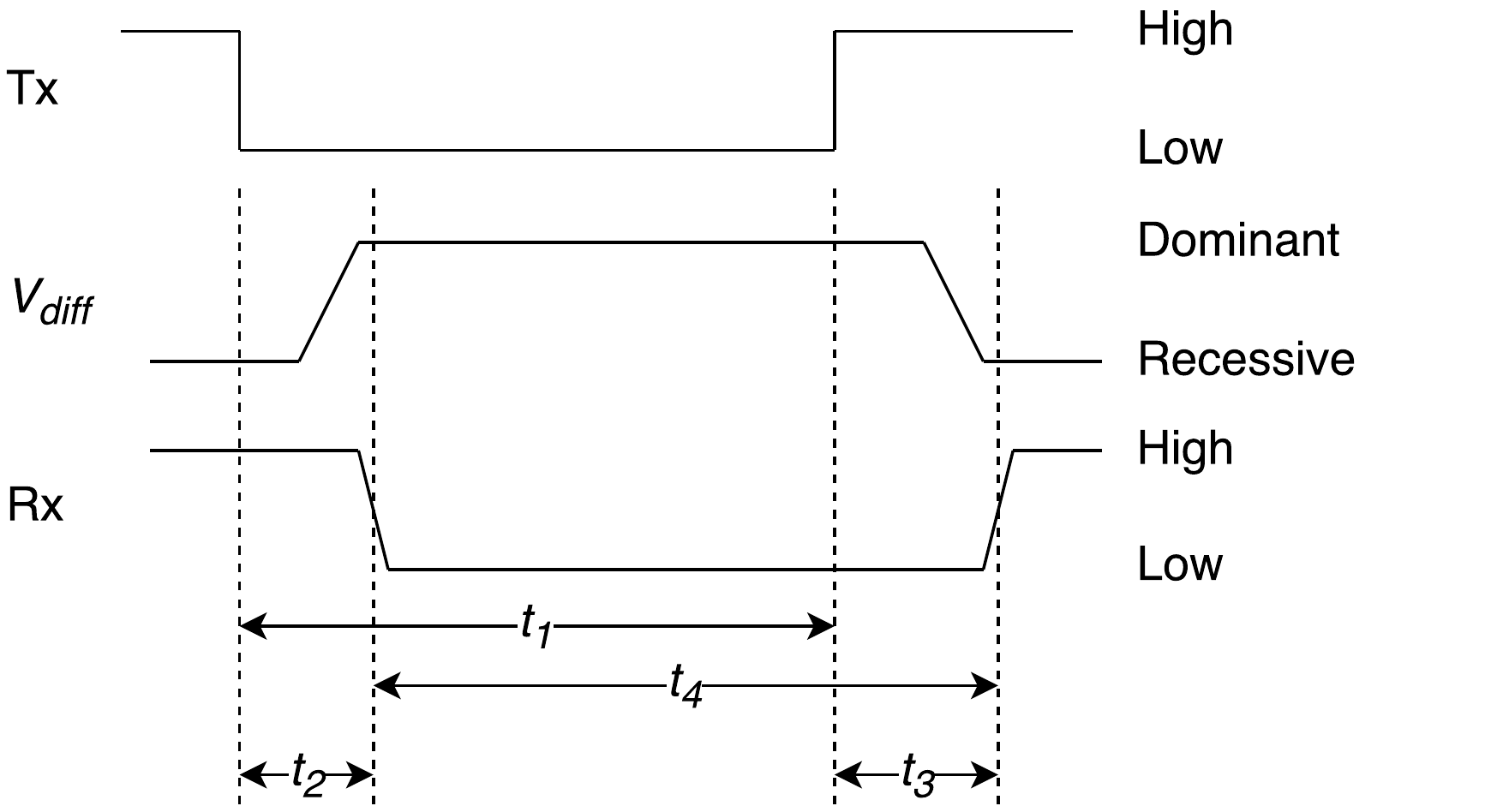}
   \caption{Delay model in CAN.}
   \label{fig:CAN_timing_diagram}
\end{figure}
As in the environment shown in Fig. \ref{fig:trans_delay} \subref{2_a}, the probe cannot be installed on the ECU's Tx to be identified on a real vehicle. Therefore, the method calculates the delay-time using only the information that can be observed by the receiving node.
Fig. \ref{fig:CAN_timing_diagram} shows a timing chart of CAN bus signal and transceiver on CAN bus.
There is a delay-time until the differential voltage ($V_{diff}$) is generated after the Tx pin of the CAN transceiver of the sending node changes and the Rx pin of the CAN transceiver of the receiving node changes.
We specified the bit time period which the sender node generates as $t_1$, the delay-time between Tx pin signal on sender CAN controller falls and Rx pin signal on receiver CAN transceiver falls as $t_2$, the delay-time between Tx pin signal when sender rises and Rx pin signal when receiver rises as $t_3$, and the bit time period receiver CAN transceiver observes as $t_4$.
$t_4$ can be represented as:
\begin{equation}
t_4 = t_1 + t_3 - t_2
\end{equation}

At this point, the time actually measured at the receiver node is only $t_4$, and $t_1, t_2, t_3$ are unknown.
Therefore, we consider the relationship between $t_1$ and $t_{bit}$.
Every CAN node has an independent clock source and communicates while synchronizing every falling edge.
Hence, we define error of crystal oscillator as $t_{e}$. The relation with $t_1$ is represented as:
\begin{equation}
t_1 = t_{bit} + t_e
\end{equation}
A crystal oscillator is used in high-speed CAN node to satisfy the frequency tolerance requirement.
A typical crystal oscillator has a frequency tolerance of about $\SI{\pm30}{ppm\, }$ to $\SI{\pm100}{ppm\, }$ \cite{murata}.
We define frequency tolerance as $f_{tol}$ [ppm] and frequency deviation $\varDelta f_e$ and the error of crystal oscillator $t_e$ are represented as follows:
\begin{eqnarray}
\varDelta f_e = f_{bit} \times f_{tol} \times 10^6 \SI{}{\ Hz}\\
t_e = \frac{1}{f_{bit} + \varDelta f_e} - t_{bit}
\end{eqnarray}

For instance,
if we assume when CAN bus baudrate is set to \SI{500}{\kilo bps} and frequency deviation is larger, $f_{bit} = \SI{500}{\kilo Hz}, f_{tol} = + \SI{100}{ppm}$.
$\varDelta f_e$ and $t_e$ are:
\begin{eqnarray}
\varDelta f_e = (500 \times 10^{3}) \times (100 \times 10^{-6}) & = & \SI{50}{Hz} \\
t_e = \frac{1}{500 \times 10^{3} + 50} - \frac{1}{500 \times 10^{3}} & \approx & -1.9998 \times 10^{-10} \, \mathrm{s} \nonumber \\
\end{eqnarray}
$-1.9998 \times 10^{-10} \, \mathrm{s}$ equals $\SI{-0.19998}{\nano s}$ and
$t_3-t_2$ is of the order of $\SI{\pm80}{ns}$.
From this, $t_{e}$ is sufficiently smaller than $t_3-t_2$.
Hence, we consider $t_1 = t_{bit}$ then $t_3 - t_2$ can be approximated:
\begin{equation}
t_3 - t_2 \approx t_4 - t_{bit}
\end{equation}

\subsubsection{Measurement period of delay-time}
\label{sec:meas}
As we showed in Fig. \ref{fig:CAN_frame}, length of a CAN data frame is variable and it is set in the DLC field. 
Therefore, even if the length of the CAN frame is the shortest (DLC=0), it is necessary to reliably be able to measure the section transmitted by the target node.
Then, considering CAN frame such as DLC=0,
\SI{35}{bits} of signal of
SOF (\SI{1}{bit}),
the arbitration field (\SI{12}{bits}),
the control field (\SI{6}{bits}) and
CRC filed (\SI{16}{bits})
are transmitted 
by the ACK field.
Here, if we include the CRC delimiter to the measurement period, there is a possibility that the rising edge of the ACK slot is measured.
We subtract \SI{1}{bit} from 35.
Hence, we set the measurement period from SOF to time that passing 34 bits time (\SI{68}{\micro s}).
Since the length of 1 frame never be shorter than the CAN frame when DLC=0, this allows us to reliably measure only the signal of the target node. Also, during the measurement of delay-time, the time capture is performed every rising edge of the Rx pin.

We describe how to obtain delay-time, $t_{delay}$ from the measured counter value.
As the unit of timer counter value is \SI{20}{\nano s},
The elapsed time from SOF, $t_{elapsed}\, \mathrm{(ns)}$ can be calculated as:
\begin{equation}
t_{elapsed} = (\mathrm{capture\, counter\, value} - \mathrm{SOF\, counter\, value}) \times 20
\end{equation}

The value of elapsed bits from the SOF at each rising edge can be calculated as follows:
\begin{equation}
\lfloor\frac{t_{elapsed}+500}{2000}\rfloor
\end{equation}
where, 500 is added in the numerator to round $t_{elapsed}$ by \SI{1000}{\nano s}, 2000 is the value of $t_{bit}$ in \SI{}{\nano s}. Also, 500 is offset to obtain the correct elapsed bits. And the ideal value of elapsed bits can be obtained with floor function.

Therefore, the ideal elapsed time from SOF, $t_{ideal}\, \mathrm{(ns)}$ can be calculated as follows:
\begin{equation}
t_{ideal}= \lfloor\frac{t_{elapsed}+500}{2000}\rfloor \times 2000
\end{equation}
$t_{delay}\, \mathrm{(ns)}$ we want to calculate is:
\begin{equation}
t_{delay} = t_{elapsed} - t_{ideal} 
= t_{elapsed} - \lfloor\frac{t_{elapsed}+500}{2000}\rfloor \times 2000
\label{eq:t_delay}
\end{equation}

\subsection{Feature Extraction}
In order to efficiently classify ECUs, we select the suitable statistical features. Similar to the conventional method \cite{kneib2018scission}, we select the features from several statistical characteristics (see Table \ref{table:stat_features}). 
\begin{table}[tbp]
\centering
\caption{A list of statistical features considered in the selection. $x$ is the delay-time in one CAN message, $N$ is the number of measured delay-time in one CAN message. }
\label{table:stat_features}
\begin{tabular}{c|c} \hline
{\bf Feature} & {\bf Description} \\ \hline
Mean & $\mu = \frac{1}{N} \sum_{i=1}^{N} x(i)$ \\
Standard Deviation & $\sigma = \sqrt{\frac{1}{N} \sum_{i=1}^{N} (x(i)-\mu)^2}$ \\
Variance & $\sigma^2 = \frac{1}{N} \sum_{i=1}^{N} (x(i)-\mu)^2$ \\
Skewness & $skew = \frac{1}{N} \sum_{i=1}^{N} (\frac{x(i)-\mu}{\sigma})^3$ \\
Kurtosis & $kurt = \frac{1}{N} \sum_{i=1}^{N} (\frac{x(i)-\mu}{\sigma})^4$ \\
Root Mean Square & $rms = \sqrt{\frac{1}{N} \sum_{i=1}^{N} x(i)^2}$ \\
Max & $max = $max$(x(i))_{i=1...N}$ \\
Energy & $en = \frac{1}{N} \sum_{i=1}^{N} x(i)^2$ \\ \hline
\end{tabular}
\end{table}
Also, We have used the Relief-F algorithm \cite{kononenko1994relief} to select the suitable features using a weight of each feature from the Weka 3 Toolkit \cite{smith2016weka}. 
The Relief-F is a filter method. It is possible to rank and select the most significant features. We conducted Relief-F to the delay-time from prototype and real-vehicle. 
As a result, the Relief-F calculated the features of ranking as shown in Table \ref{table:rank_features}. 
\begin{table}[tbp]
\centering
\caption{Ranking of the features calculated by Relief-F algorithm \cite{kononenko1994relief}. }
\label{table:rank_features}
\begin{tabular}{r|cc|cc} \hline
   & {\bf Prototype}    & {\bf Weight} & {\bf Real-vehicle} & {\bf Weight} \\ \hline
1. & Mean               & 0.1195 & Root Mean Square   & 0.2401 \\
2. & Root Mean Square   & 0.1060 & Max                & 0.2001 \\
3. & Max                & 0.0467 & Mean               & 0.1910 \\
4. & Standard Deviation & 0.0435 & Energy             & 0.1792 \\
5. & Energy             & 0.0314 & Kurtosis           & 0.0934 \\
6. & Kurtosis           & 0.0310 & Skewness           & 0.0692 \\
7. & Skewness           & 0.0220 & Standard Deviation & 0.0250 \\
8. & Variance           & 0.0125 & Variance           & 0.0177 \\ \hline
\end{tabular}
\end{table}
In both prototype and real-vehicle, we confirmed that Mean, Root Mean Square, and Max are ranked from 1st to 3rd in the ranking. This result suggests that we can efficiently classify ECUs either prototype or real-vehicle using the selected features. Hence, we use Mean, Root Mean Square, and Max as the suitable statistical features to efficiently classify ECUs.

\subsection{Classification}
The sender identification can result in a classification problem. We use $k$-nearest neighbor ($k=5$) in Divider. The algorithm is the simplest in machine learning algorithms. In addition, it is possible that ECUs' limited resources in the in-vehicle system can execute $k$-nearest neighbor. 

\subsection{Implementation}
\begin{figure}[tbp]
  \centering
  \begin{minipage}{0.8\hsize}
   \centering
   \includegraphics[width=1.0\textwidth]{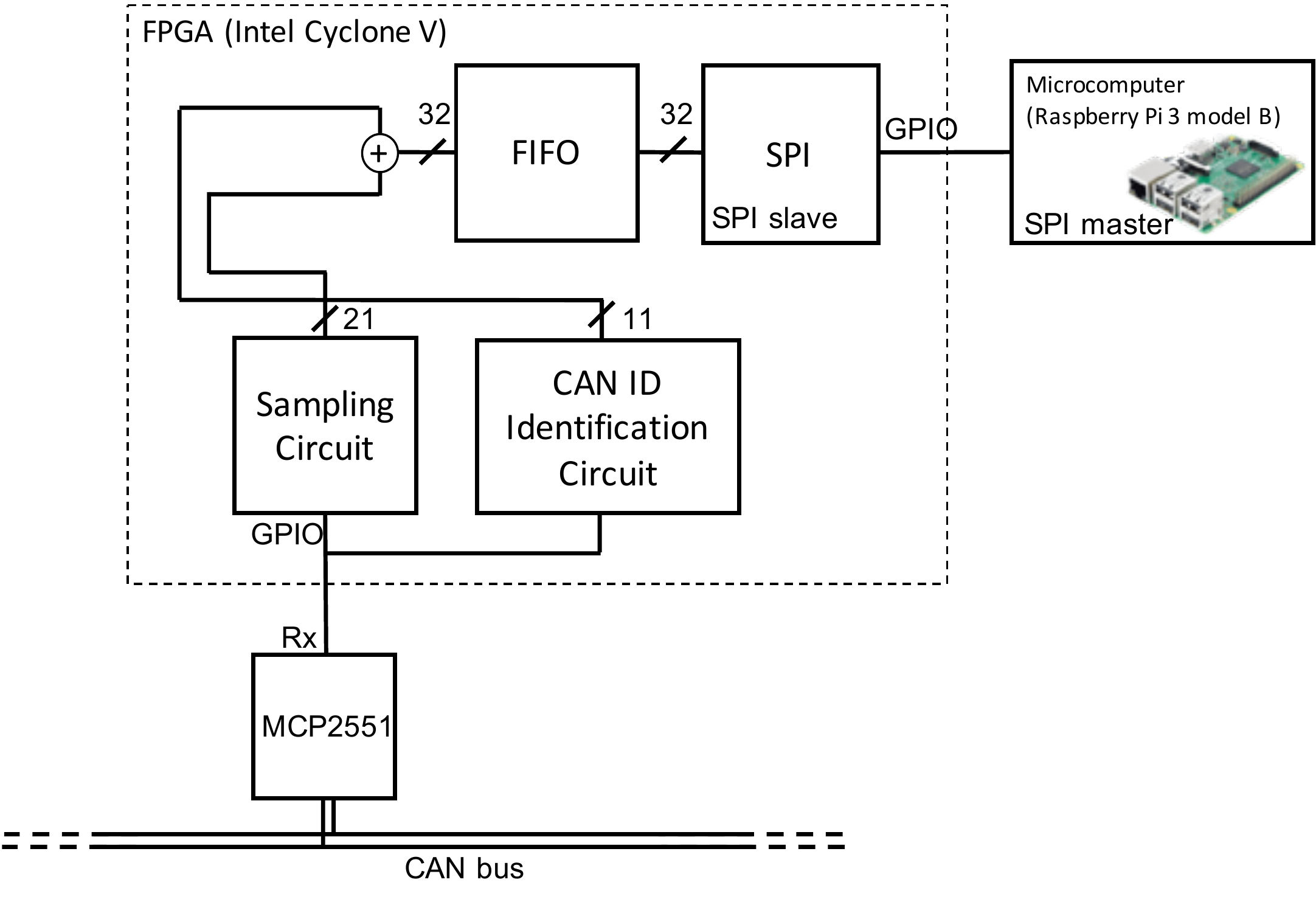}
   \subcaption{Implementation of proposed method.}\label{fig:implement}
  \end{minipage}
  \begin{minipage}{0.7\hsize}
  \centering
   \includegraphics[width=1.0\textwidth]{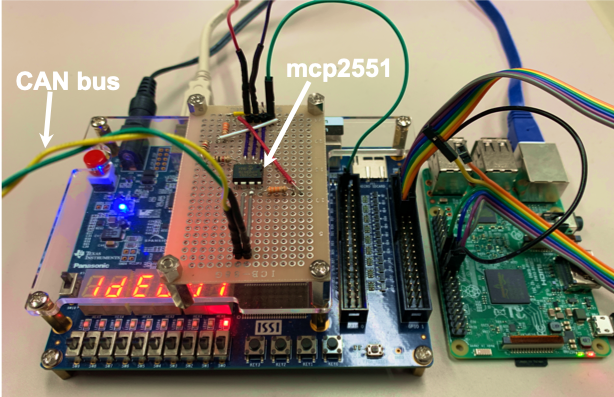}
   \subcaption{Prototype of IDS using proposed method.}\label{fig:ids_pic}
   \end{minipage}
   \caption{Implementation and prototype.}
   \label{fig:implementation_and_prototype}
\end{figure}

In this section, we describe the implementation of Divider. As mentioned in Section \ref{sec:meas}, the proposed method measures the 34 bits to observe delay-time no matter what length of the data field is received. We show the block diagrams of the implementation of Divider in Fig. \ref{fig:implementation_and_prototype} \subref{fig:implement}.
The MCP2551 is a chip that serves CAN transceiver as the interface between a CAN controller and the physical bus. We also selected a Field Programmable Gate Array (FPGA) as a measurement device, because a measurement with software cannot process all messages without missing ones due to the limitation of the ability of microcomputer. We show the prototype of IDS in Fig. \ref{fig:implementation_and_prototype} \subref{fig:ids_pic}. We developed the prototype of the proposed method using FPGA and microcomputer. We selected the DE0-CV Cyclone V Board (5CEBA4F23C7) as FPGA and Raspberry Pi 3 model B as microcomputer. Also, we release the source code \cite{Divider} of the proposed method written by Verilog and Python hoping to promote the research of the sender identification method.

Here, we describe the circuits of FPGA in the proposed method. The circuits are divided into four operations. The first is sampling circuit. This circuit measures a period of CAN message in the measurement period of 34 bits with counting. The sampling circuit sends counting value to FIFO per Rx rising edge of CAN. The second is the arbitration ID identification circuit. As its name suggests, we observe and store the arbitration ID of every message. Like the sampling circuit, the arbitration ID sampling circuit sends arbitration ID to FIFO per Rx rising edge too. The third is FIFO. Here, we stack the measured data of 32 bits constructed from an arbitration ID of 11 bits and the counter value of the dominant period of 21 bits. The last is the SPI module. We implement the SPI slave module to send measurement data to the Raspberry Pi. 

The operation of the measurement is as follows. 
\begin{enumerate}
 \item Starting the capture of measurement time and arbitration ID, an occurrence at the falling edge of SOF bit.
 \item Send the measurement data (arbitration ID and measurement time) with every rising edge of Rx to FIFO. 
       Also, after Raspberry Pi receives the measurement data from the FPGA, calculate the delay-time by equation (\ref{eq:t_delay}) and record the delay-time.
 \item After \SI{34}{bits} from SOF, the measurement is ended.
 \item When CAN frame is completely received, the sampling circuit and arbitration ID identification circuit are waiting SOF bit.
\end{enumerate}

\section{Evaluation}
\label{section:eval}
\subsection{Environments}
In this section, we evaluate Divider of the proposed method on a prototype of CAN bus, and real-vehicle. 

\begin{figure}[tbp]
  \centering
  \begin{minipage}{1.0\hsize}
    \centering
    \includegraphics[width=1.0\textwidth]{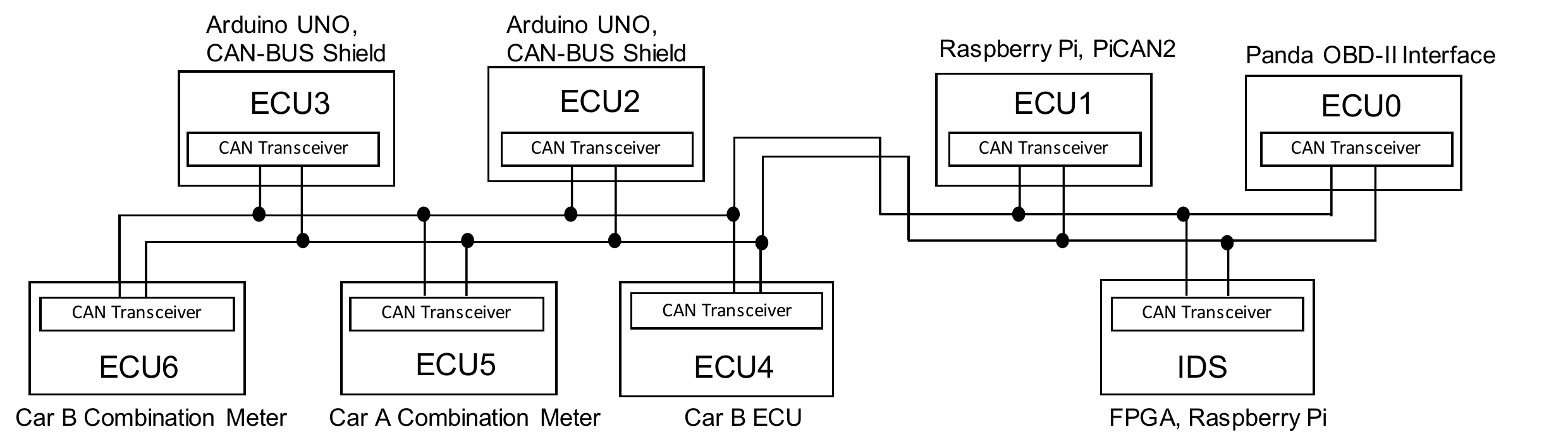}
    \subcaption{The prototype of CAN bus.}
    \label{fig:proto}
  \end{minipage}
  \begin{minipage}{1.0\hsize}
    \centering
    \includegraphics[width=1.0\textwidth]{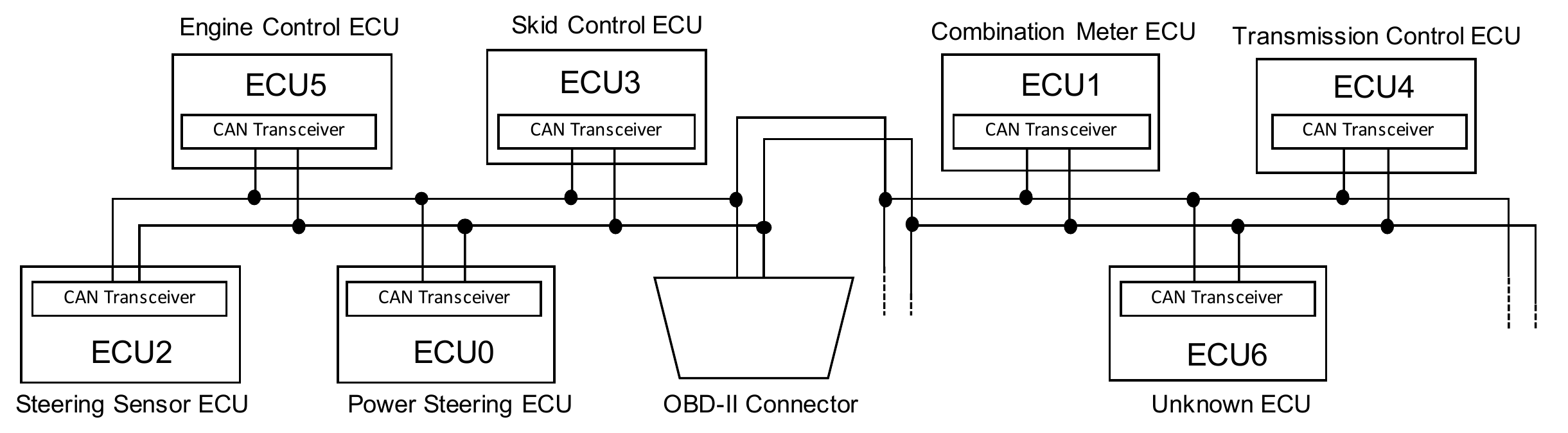}
    \subcaption{A part of CAN bus in our real-vehicle.}
    \label{fig:vehicle}
  \end{minipage}
  \caption{Environments.}
  \label{fig:proto_vehicle}
\end{figure}

Fig. \ref{fig:proto_vehicle} \subref{fig:proto} shows the prototype of the CAN bus topology we implemented in our experiment. 
We prepare various ECUs to evaluate Divider. 
The various ECUs we prepared are described here. ECU0 is panda OBD-II interface \cite{Panda}, ECU1 is Raspberry Pi model B mounted with PiCAN 2 board, ECUs 2 and 3 are Arduino UNO mounted with CAN-BUS Shield, ECU4 is an actual ECU not connected other than CAN, ECUs 5 and 6 are an actual combination meter of each different car model. We cannot control sending CAN messages of ECUs 4, 5, and 6 but these ECUs automatically send some messages periodically, so that Divider uses the messages to fingerprint ECU.

Fig. \ref{fig:proto_vehicle} \subref{fig:vehicle} shows a part of CAN in real-vehicle which is used to evaluate Divider. 
The real-vehicle has multiple CAN buses. 
One of these CAN buses has a realistic environment in which each ECU has a yaw-rate sensor or an acceleration sensor sends the information to the meter ECU. 
This CAN bus also has OBD-II port. 
In the real-vehicle experiment, we have collected the datasets during driving and stopping.

First, we show the ability to fingerprint various ECUs in Divider. Next, we evaluate the accuracy of intrusion detection of Divider against a compromised ECU and an unmonitored ECU.

\subsection{Identification of various ECU: Prototype and real-vehicle}
\subsubsection{Prototype of CAN bus}
First of all, we evaluate the ability to the identification of various ECU in the prototype environment. The bus topology of the prototype is shown in Fig. \ref{fig:proto_vehicle} \subref{fig:proto}. 
We have captured 1000 messages from each ECU. The 1000 messages are used to calculate the features of Mean, Root Mean Square, and Max. We use the features to learn sender characteristics. Next, we divided the messages into learning data (\SI{80}{\%}) and testing data (\SI{20}{\%}). Hence, we evaluate the proposed method using $K$-fold cross validation in $K=5$.

\begin{figure}[tbp]
  \centering
  \begin{minipage}{0.8\hsize}
    \centering
    \includegraphics[width=1.0\textwidth]{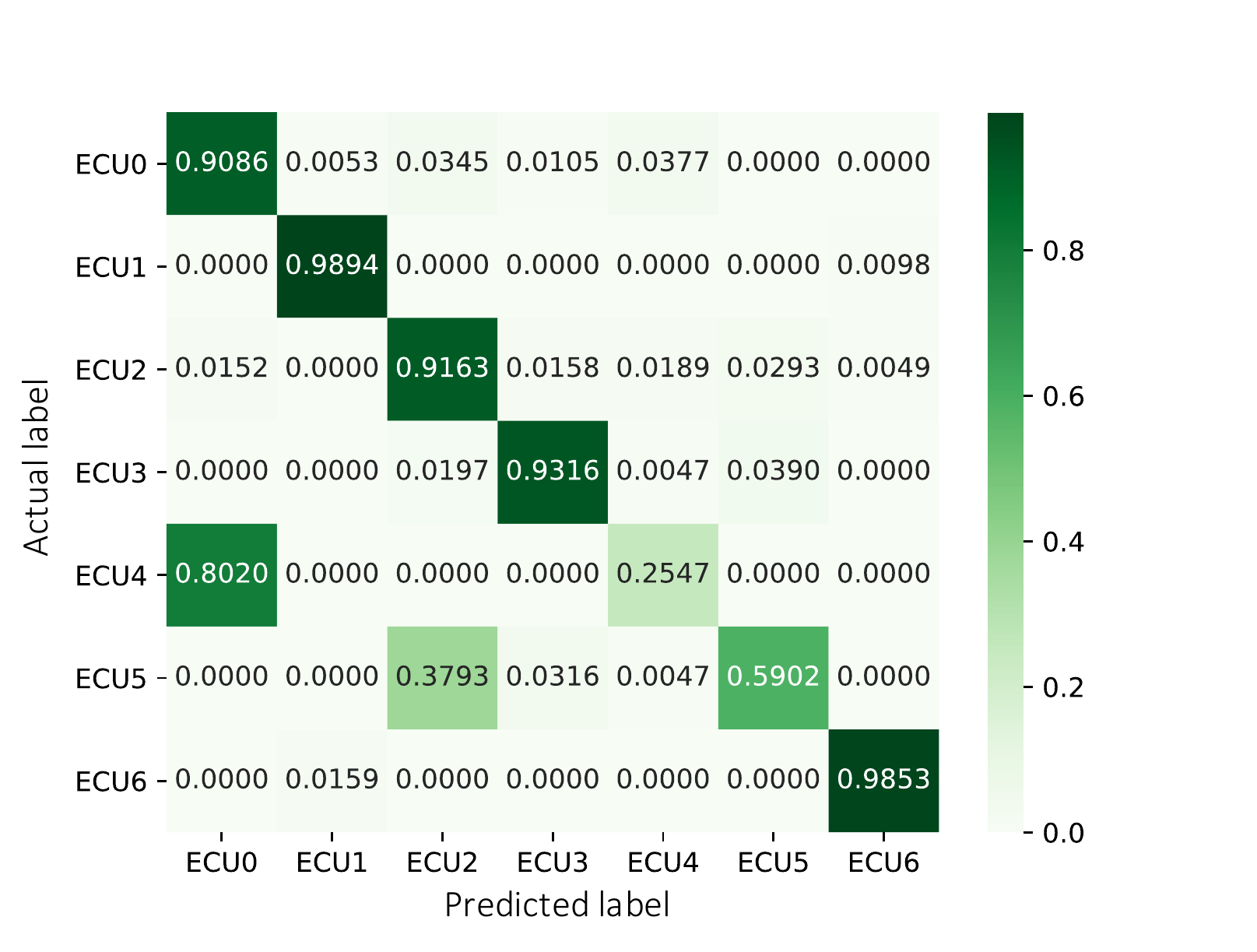}
    \subcaption{The prototype.}
    \label{fig:knn_confusion_proto}
  \end{minipage}
  \begin{minipage}{0.8\hsize}
    \centering
    \includegraphics[width=1.0\textwidth]{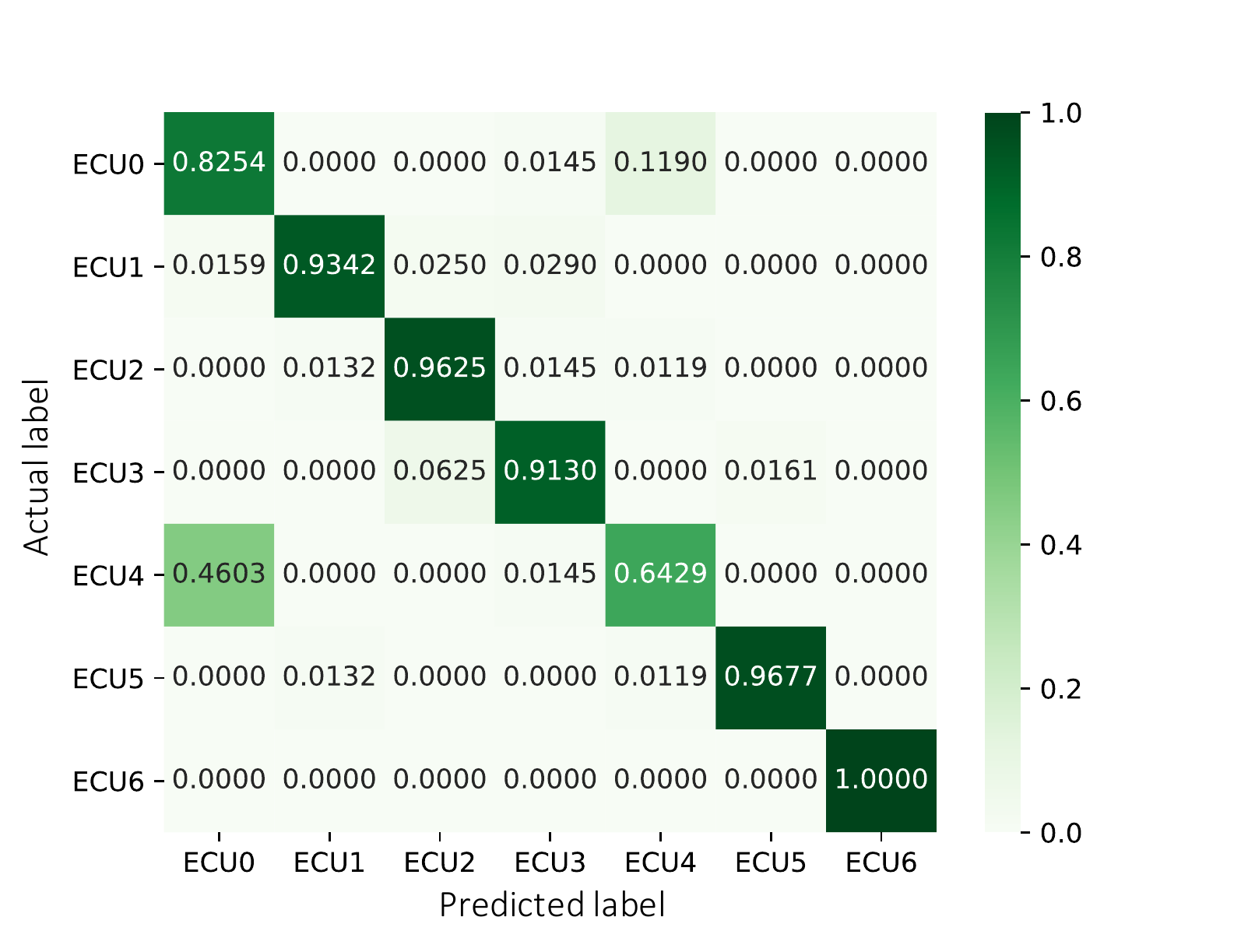}
    \subcaption{The real-vehicle.}
    \label{fig:knn_confusion_prius}
  \end{minipage}
  \caption{Confusion matrix for the identification of ECUs.}
  \label{fig:knn_confusion}
\end{figure}

As a result, an average of accuracy is \SI{79.07}{\%}. The one confusion matrix in $K$-fold cross validation is shown in Fig. \ref{fig:knn_confusion} \subref{fig:knn_confusion_proto}. 
It can be seen that Divider can identify correctly with up to \SI{98.94}{\%}. While a minimal identification rate is \SI{25.47}{\%}. 

\subsubsection{Real-vehicle}
We have also evaluated ECU identification accuracy in real vehicle's CAN bus.
We have captured 200000 messages from the ECUs. We extracted the feature data from the 360 messages of each ECU to align the number of messages of each ECU in the learning and verification data. As with the prototype, we divided the data of the delay-time of each arbitration ID into learning data (\SI{80}{\%}) and testing data (\SI{20}{\%}). Hence, we evaluate the proposed method using $K$-fold cross validation in $K=5$.

From the $K$-fold cross validation, Divider performed well with an average accuracy of \SI{88.77}{\%}.
The confusion matrix is shown in Fig. \ref{fig:knn_confusion} \subref{fig:knn_confusion_prius}. It can be seen that Divider can identify correctly with up to \SI{100.00}{\%}. While a minimal identification rate is \SI{64.29}{\%}. 

\subsection{Intrusion Detection}
\subsubsection{Compromised ECU}
In this section, we evaluate the intrusion detection capability of the learned model. To reproduce compromised ECU, we sent an arbitration ID: $x$ assigned in ECU6 from ECU1 spoofed to ECU6. Spoofing attacks were performed for three minutes from ECU1, and the data during attacks of ECUs 1 and 6 were classified by the learned model. The results are shown in Table \ref{table:eval_proto}. Predicted: Attack label is when Divider classifies messages of ID: $x$ as other than ECU6, Predicted: Normal label is when Divider classifies messages of ID: $x$ as ECU6. We confirm the true positive rate against compromised ECU is \SI{96.7}{\%} and the true negative rate is \SI{83.2}{\%}. 
\begin{table}[tbp]
\centering
\caption{Confusion matrix against sending ID: $x$ from compromised ECU (ECU1) spoofed to ECU6.}
\label{table:eval_proto}
\begin{tabular}{c|c|c} \hline
 & Predicted: Attack & Predicted: Normal \\ \hline
Actual: Attack & 0.967 & 0.033 \\
Actual: Normal & 0.168 & 0.832 \\ \hline
\end{tabular}
\end{table}
\subsubsection{Unmonitored ECU}
Similar to compromised ECU, we evaluate the ability of intrusion detection against unmonitored ECU. We attached the Arduino UNO (the ECU2 in the prototype of CAN bus) as an unmonitored ECU in CAN of real-vehicle. We assume the spoofing attacks of speed information from the unmonitored ECU. Therefore, the unmonitored ECU sends ID: $y$ assigned as arbitration ID of speed in real-vehicle. Spoofing attacks were performed for three minutes from unmonitored ECU, and the data during sending messages of ECU3 (legitimate ECU of ID: $y$) and unmonitored ECU were classified by the learned model. The results are shown in Table. \ref{table:eval_prius}. 
We confirm the true positive rate against unmonitored ECU is \SI{98.0}{\%} and the true negative rate is \SI{92.0}{\%}.

\begin{table}[tbp]
\centering
\caption{Confusion matrix against sending ID: $y$ from unmonitored ECU spoofed to ECU3.}
\label{table:eval_prius}
\begin{tabular}{c|c|c} \hline
 & Predicted: Attack & Predicted: Normal \\ \hline
Actual: Attack & 0.980 & 0.020 \\
Actual: Normal & 0.080 & 0.920 \\ \hline
\end{tabular}
\end{table}

\section{Discussions}
\label{section:discussions}
\subsection{Fingerprinting ECUs} 
As Fig. \ref{fig:knn_confusion} \subref{fig:knn_confusion_prius} shows, the number of ECUs with a classification accuracy of more than 80\% are 6 out of 7 in the real-vehicle. Thus, our evaluations clearly showed the difference in the delay-time of some ECUs. 

While in the prototype, a minimal identification rate is 25.47\%. 
This lower result was caused by the difference of delay-time is sometimes close between different ECUs. 

In such cases, the proposed method cannot classify the ECUs correctly, because the experimental device does not have sufficient time resolution (\SI{20}{ns}). 
Since a device of sufficient time resolution can more sparsely divide the delay-times, the proposed method using the device of sufficient time resolution will classify ECUs with high accuracy more than our experimental device. 
Therefore, we consider improving the time resolution as future work.

\subsection{Number of sampling}
Next, we discuss the number of samplings performed by sender identification methods for each CAN message. 
Table \ref{table:comp_conventional} shows a comparison among the methods.
The number of samplings per CAN message for Choi's method, Scission, and SIMPLE depends on the length of the data field. Thus, we consider the case when the data field is the shortest (\SI{0}{byte}) and longest (\SI{8}{byte}). If the data field is shortest (\SI{0}{byte}), the length of CAN message is \SI{47}{bit} from Fig. \ref{fig:CAN_frame}. Also, when the bit rate of CAN is \SI{500}{kbps}, the transmission time for \SI{1}{bit} is \SI{2}{\micro \second}. Hence, the sampling rate of each method is multiplied by $47 \times 2 \times 10^{-6}$. As a result, the best number of sampling per CAN message is $198 \times 10^{3}$, $1980$, $47$ respectively. Similarly, if the data field is longest (\SI{8}{byte}), the length of CAN message is \SI{111}{bit}. Therefore, the worst number of sampling per CAN message is $444 \times 10^{3}$, $4440$, $111$ respectively. The number of samplings per message in the proposed method depends on the number of signal transitions from 0 to 1, not the length of the data field. 
Consequently, The minimum and the maximum number of sampling of the proposed method are discussed with Arbitration ID 0x000, which has a small number of bit transitions, and Arbitration ID 0x555, which has a large number of transitions. As a result, in the case of arbitration ID 0x000, the best number of sampling reached 5. In the case of arbitration ID 0x555, the worst number of sampling reached 14. The results show that the proposed method has the least number of samplings at the data acquisition phase; in other words, the proposed method has the smallest $n$ at the feature extraction stage. Hence, the feature extraction of Divider is possible with light processing.

Finally, we discuss computational complexity. The method of Choi et al. and Scission use time and frequency domain features. Therefore, these methods need $\mathrm{\Omega}(n \log n)$ time because these methods perform Fourier Transforms to calculate the frequency domain feature. Also, since SIMPLE calculates the mean as a statistic with a time domain feature, it takes $\mathrm{\Theta}(n)$. Similarly, because Divider uses statistic features in Table. \ref{table:stat_features}, Divider needs $\mathrm{\Theta}(n)$. Therefore, we confirmed that the computational complexities of SIMPLE and Divider are lower than the computational complexities of other methods.

From these comparisons among related works, we confirmed that Divider can reduce the amount of data in the data acquisition phase than the other voltage-based methods. 

\begin{table}[tbp]
\centering
    \caption{Comparison among sender identification methods in Sampling Rate (S.R.), Best Number of Sampling per message (B.N.S.), Worst Number of Sampling per message (W.N.S.), Time Complexity (T.C.).}
\label{table:comp_conventional}
\begin{tabular}{r|r|r|r|r} \hline
         & Choi et al. \cite{choi2018identifying} & Scission \cite{kneib2018scission} & SIMPLE \cite{foruhandeh2019simple} & Divider \\ \hline
    S.R. & \SI{2}{GS/s} & \SI{20}{MS/s} & \SI{500}{kS/s} & - \\
    B.N.S. & $198 \times 10^{3}$ & $1980$ & $47$ & 5 \\
    W.N.S. & $444 \times 10^{3}$ & $4440$ & $111$ & 14 \\
    T.C. & $\mathrm{\Omega}(n \log n)$ & $\mathrm{\Omega}(n \log n)$ & $\mathrm{\Theta}(n)$ & $\mathrm{\Theta}(n)$ \\ \hline
\end{tabular}
\end{table}

\section{Conclusions}
To avoid the security risk on automobiles, IDSs using features of physical-level such as the voltage value has been proposed. 
However, these IDSs require high sampling rates and high computing resources.
In this research, we proposed a delay-time based sender identification method which is low sampling rate. We implemented the experimental devices using FPGA and microcomputer to verify our method for identification. As a result, we confirm that the proposed method achieved a true positive rate of \SI{96.7}{\%} in CAN bus prototype against spoofing attack from compromised ECU. We have released our research \cite{Divider} in the hope to promote research on sender identification. 
In our future work, we plan to improve the time resolution of Divider and to try various learning algorithms such as random forest classifier. Furthermore, we will consider the Intrusion Prevention System based on the sender identification method.

\bibliographystyle{IEEEtran}
\bibliography{bibfile}

\end{document}